\documentclass[aps,prl,twocolumn,showpacs,preprintnumbers,amsmath,amssymb,superscriptaddress]{revtex4}%

\usepackage{graphicx}%
\usepackage{dcolumn}
\usepackage{amsmath}
\usepackage{color}
\usepackage{bm}

\begin{document}

\preprint{PREPRINT (\today)}

\title{Oxygen isotope effect on superconductivity and magnetism
in Y$_{1-x}$Pr$_{x}$Ba$_2$Cu$_3$O$_{7-\delta}$.}

\author{R.~Khasanov}
 \affiliation{Physik-Institut der Universit\"{a}t Z\"{u}rich,
Winterthurerstrasse 190, CH-8057 Z\"urich, Switzerland}
\author{A.~Shengelaya}
 \affiliation{Physik-Institut der Universit\"{a}t Z\"{u}rich,
Winterthurerstrasse 190, CH-8057 Z\"urich, Switzerland}
 \affiliation{Physics Institute of Tbilisi State University,
Chavchavadze 3, GE-0128 Tbilisi, Georgia}
\author{D.~Di~Castro}
 \affiliation{Physik-Institut der Universit\"{a}t Z\"{u}rich,
Winterthurerstrasse 190, CH-8057 Z\"urich, Switzerland}
 \affiliation{"Coherentia" CNR-INFM and Dipartimento di Fisica,
Universita' di Roma "La Sapienza", P.le A. Moro 2, I-00185 Roma,
Italy}
\author{E.~Morenzoni}
 \affiliation{Laboratory for Muon Spin Spectroscopy, Paul Scherrer
Institut, CH-5232 Villigen PSI, Switzerland}
\author{A.~Maisuradze}
 \affiliation{Physik-Institut der Universit\"{a}t Z\"{u}rich,
Winterthurerstrasse 190, CH-8057 Z\"urich, Switzerland}
\author{I.M.~Savi\'c}
 \affiliation{Faculty of Physics, University of Belgrade, 11001
Belgrade, Serbia and Montenegro}
\author{K.~Conder}
 \affiliation{Laboratory for Developments and Methods, Paul Scherrer Institute,
CH-5232 Villigen PSI, Switzerland}
\author{E.~Pomjakushina}
 \affiliation{Laboratory for Developments and Methods, Paul Scherrer Institute,
CH-5232 Villigen PSI, Switzerland}
 \affiliation{Laboratory for Neutron Scattering, Paul Scherrer Institute \& ETH
 Zurich,CH-5232 Villigen PSI, Switzerland}
\author{H.~Keller}
\affiliation{Physik-Institut der Universit\"{a}t Z\"{u}rich,
Winterthurerstrasse 190, CH-8057 Z\"urich, Switzerland}
%

\begin{abstract}
Oxygen isotope ($^{16}$O/$^{18}$O) effects (OIE's) on the
superconducting transition ($T_c$), the spin-glass ordering
($T_g$), and the antiferromagnetic ordering ($T_N$) temperatures
were studied for Y$_{1-x}$Pr$_x$Ba$_2$Cu$_3$O$_{7-\delta}$  as a
function of Pr content ($0.0 \leq x \leq 1.0$). The OIE on $T_c$
increases with increasing $x$ up to $x \approx 0.55$, where
superconductivity disappears. For decreasing $x$ the OIE's on
$T_N$ and $T_g$ increase down to  $x\approx 0.7$ where
antiferromagnetic order and down to $x\approx 0.3$ where
spin-glass behavior vanish, respectively. The OIE's on $T_g$ and
$T_N$ are found to have {\it opposite signs} as compared to the
OIE on $T_c$. All OIE's are suggested to arise from the isotope
dependent mobility (kinetic energy) of the charge carriers.

\end{abstract}

\maketitle
\begin{figure}[htb]
\includegraphics[width=0.7\linewidth, angle=90]{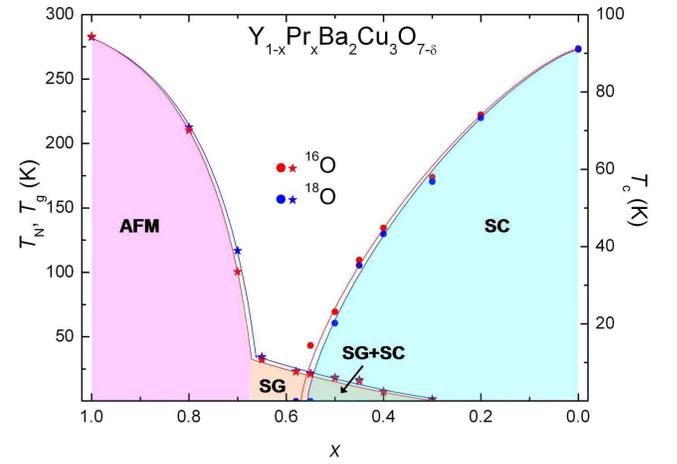}
 \vspace{-0.3cm}
\caption{ Dependence of the superconducting transition ($T_c$),
the spin-glass ordering ($T_g$), and the antiferromegnetic
ordering ($T_N$) temperatures for $^{16}$O/$^{18}$O substituted
Y$_{x}$Pr$_{1-x}$Ba$_2$Cu$_3$O$_{7-\delta}$ on the Pr content $x$.
The solid lines for $^{16}T_g(x) \, [^{18}T_g(x)]$ and
$^{16}T_N(x) \,[^{18}T_g(x)]$  are guides to the eye and those for
$^{16}T_c(x) \, [^{18}T_c(x)]$ result from the power law fit (see
text for details). The areas denoted by ''AFM``, ''SG``, and
''SC`` represent the antiferromagnetic, the spin-glass and the
superconducting regions, respectively. ''SG+SC`` corresponds to
the region where spin-glass magnetism coexist with
superconductivity. }
 \label{fig:Phase-Diagramm}
\end{figure}

 High-temperature cuprate superconductors (HTS's) exhibit
a rich phase diagram as a function of doping (see {\it e.g.}
Fig.~\ref{fig:Phase-Diagramm}). The undoped parent compounds are
characterized by a long range 3D antiferromagnetic (AFM) order which
is rapidly destroyed when holes are doped into the CuO$_2$ planes.
The short-range AFM correlations survive, however, well in the
superconducting (SC) region of the phase diagram by forming a
spin-glass (SG) state. The issues of the interplay of magnetism and
superconductivity in HTS's and the nature of doping-induced charge
carriers within the antiferromagnetic CuO$_2$ planes are still
controversial. Understanding these fundamental questions can help to
clarify the pairing mechanism of high-temperature superconductivity.

In conventional superconductors, key experimental evidence for a
phonon mediated pairing mechanism was provided by measurements of
the isotope effect on the transition temperature $T_c$. In
contrast, in HTS's a number of unconventional oxygen-isotope
($^{16}$O/$^{18}$O)  effects (OIE's) on various physical
quantities, including among others the transition temperature
$T_c$, the in-plane magnetic penetration depth $\lambda_{ab}(0)$,
the pseudogap temperature $T^\ast$, and the spin-glass temperature
$T_g$, were observed which cannot be explained by standard BCS
theory
\cite{Batlogg87a,Franck91Franck94,Zech94,Zhao97Zhao98,Zhao01,
Hofer00,Khasanov04a,Khasanov04Khasanov06Khasanov07,Keller05,Tallon05,
Zhao94,Shengelaya99}. For instance, it was found that the OIE's on
$T_c$ and $\lambda_{ab}(0)$ are strongly doping dependent
\cite{Batlogg87a,Franck91Franck94,Zech94,Zhao97Zhao98,Zhao01,
Hofer00,Khasanov04a,Khasanov04Khasanov06Khasanov07,Keller05,Tallon05}.
In particular, close to optimal doping the OIE on $T_c$ is almost
zero \cite{Batlogg87a,Franck91Franck94,Zech94}, while the OIE on
$\lambda_{ab}(0)$ is still substantial
\cite{Khasanov04a,Khasanov04Khasanov06Khasanov07,Keller05,Tallon05}.
With decreasing doping both OIE's on $T_c$ and $\lambda_{ab}(0)$
increase and, for highly underdoped materials, even exceed the
value of the BCS isotope exponent $\alpha^{\rm BCS}_{T_c}=0.5$
\cite{Franck91Franck94,Zhao97Zhao98,Zhao01,Khasanov04a,Keller05,Tallon05}.
In order to obtain a more global view of cuprate superconductors a
detailed study of the isotope dependence of magnetic quantities is
needed. So far, to our knowledge only little work has been
reported on this subject. This includes experimental studies of
the OIE on the antiferromagnetic ordering temperature ($T_N$)
\cite{Zhao94} and the OIE on the spin-glass ordering temeprature
($T_g$) \cite{Shengelaya99}, as well as a theoretical
investigation of the OIE on $T_N$ \cite{Bussmann-Holder98}.
Here we report a systematic study of the OIE on $T_c$, $T_g$, and
$T_N$ in Y$_{1-x}$Pr$_x$Ba$_2$Cu$_3$O$_{7-\delta}$ as a function
of Pr content $x$ ($0.0\leq x\leq 1.0$) by means of magnetization
and muon-spin rotation ($\mu$SR) experiments. The OIE's on $T_c$
and $T_N$ were found to be vanishingly small at $x=0.0$ and
$x=1.0$, respectively, and increase for the intermediate $x$. In
the range $0.3 < x < 0.6$ where superconductivity and spin-glass
magnetism coexist both $T_c$ and $T_g$ exhibit a large OIE. The
OIE's on $T_N$ and $T_g$ are {\em sign reversed} compared to the
OIE on $T_c$.

Polycrystalline samples of Y$_{1-x}$Pr$_x$Ba$_2$Cu$_3$O$_{7-\delta}$
($0\leq x\leq 1.0$) were prepared by standard solid state reaction
\cite{Conder01}. Oxygen isotope exchange was performed during
heating the samples in $^{18}$O$_2$ gas. To ensure the same thermal
history of the substituted ($^{18}$O) and not substituted ($^{16}$O)
samples, both annealings (in $^{16}$O$_2$ and $^{18}$O$_2$ gas) were
always performed simultaneously. The $^{18}$O content in the
samples, as determined from a change of the sample weight after the
isotope exchange, was found to be 80(5)\% for all $^{18}$O
substituted samples.

\begin{figure}[htb]
\includegraphics[width=1.05\linewidth]{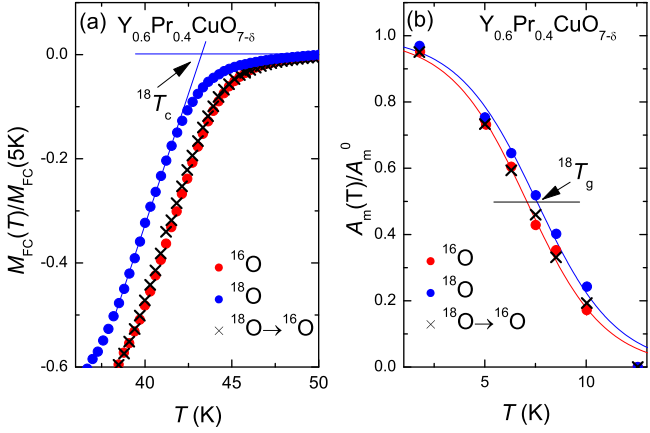}
 \vspace{-0.5cm}
\caption{ (a) Temperature dependences of the field--cooled
magnetization $M_{FC}$ normalized to its value at $T=5$~K, and (b)
$A_m/A_m^0$ for $^{16}$O, $^{18}$O, and backexchanged
$^{18}$O$\rightarrow$$^{16}$O samples of
Y$_{0.6}$Pr$_{0.4}$Ba$_2$Cu$_3$O$_{7-\delta}$. The solid lines in
(b) represent the fits of $A_m(T)$ for the $^{16}$O/$^{18}$O
substituted samples by means of Eq.~(\ref{eq:fermi}). The arrows
indicate $T_c$ and $T_g$ for the $^{18}$O substituted sample. }
 \label{fig:BE}
\end{figure}

The OIE on $T_c$ was obtained by field-cooled magnetization
($M_{FC}$) experiments performed with a SQUID magnetometer in a
field of 1.0~mT and at temperatures between $1.75$~K and $100$~K.
The values of $T_c$ were defined as the temperatures where the
linearly extrapolated $M_{FC}(T)$'s intersect the zero line [see
Fig.~\ref{fig:BE}~(a)].
The OIE's on the magnetic ordering temperatures ($T_g$ and $T_N$)
were extracted from the zero-field $\mu$SR data. No magnetism down
to $T\simeq1.7$~K was detected for the $^{16}$O/$^{18}$O substituted
samples with $x=0.0$ and $x=0.2$. For $x=0.3$ and $0.4$ magnetism
was identified as a fast decrease of the asymmetry at $T<10$~K and
$T<5$~K, respectively. For $x\geq 0.45$ damped oscillations due to
muon-spin precession in local magnetic fields were observed.  The
$\mu$SR asymmetry spectra for $x=0.8$ and $1.0$, {\it i.e.} in the
deep antiferromagnetic phase, were analyzed by using the following
expression:
\begin{eqnarray}
 A(t)&=&A_n\exp(-\sigma^2t^2/2)+A_m[\omega\exp(-\lambda_1t)
\cos(\gamma_\mu B_\mu t) \nonumber \\
 &&+(1-\omega)\exp(-\lambda_2t)J_0(\gamma_\mu
B_\mu t)].
 \label{eq:cosine-plus-bessel}
\end{eqnarray}
Here $A_m$ and $A_n$ represent the oscillating (magnetic) and
nonoscillating amplitudes, respectively, $\omega$ is a weighting
factor, $B_\mu$ is the mean internal magnetic field at the muon
site, $\gamma_{\mu} = 2\pi\times135.5342$~MHz/T is the muon
gyromagnetic ratio, and $J_0$ is the zeroth-order Bessel function.
We used the damped Bessel function $J_0$ together with the cosine
oscillating term in order to account for the unphysically large
values of the initial phase $\phi\simeq20^{\rm o}-45^{\rm o}$
which has to be introduced close to $T_N$ in order to fit the data
by using the cosine term only [$\cos(\gamma_\mu B_\mu t+\phi)$].
For $0.45\leq x\leq0.7$ and for $x=0.4$ and $0.3$ the fit was
simplified by taking from the second part of
Eq.~(\ref{eq:cosine-plus-bessel}) only the damped Bessel term and
the exponential damping term with $B_\mu=0$, respectively. The
magnetic ordering temperatures ($T_g$ and $T_N$) were then
determined by fitting the temperature dependence of $A_m$ by means
of the  phenomenological function:
\begin{equation}
A_m(T)/A_m^0=(1+\exp[(T-T_m)/\Delta T_m])^{-1}.
 \label{eq:fermi}
\end{equation}
Here $A_m^0$ is the maximum value of the asymmetry, $T_m$ is the
magnetic ordering temperature ($m=g$, $N$) and $\Delta T_m$ is the
width of the magnetic transition [see Fig.~\ref{fig:BE}~(b)].

In order to confirm the intrinsic origin of the OIE's on $T_c$ and
$T_m$, back-exchange OIE experiments were carried out for the
samples with $x=0.0$ and $x=0.4$. As shown in Fig.~\ref{fig:BE},
the $^{16}$O oxygen back exchange of the $^{18}$O sample of
Y$_{0.6}$Pr$_{0.4}$Ba$_2$Cu$_3$O$_{7-\delta}$ results within error
in almost the same $M_{FC}(T)$ [panel (a)] and $A_m(T)$ [panel
(b)] as for the $^{16}$O sample.
\begin{table*}[htb]
\caption[~]{\label{Table:OIE-results} Summary of the OIE studies
on $T_c$ and $T_m$ ($m=g$, $N$) for
Y$_{1-x}$Pr$_x$Ba$_2$Cu$_3$O$_{7-\delta}$. The meaning of the
parameters is -- $^{16}T_c$/$^{18}T_c$ and $^{16}T_m$/$^{18}T_m$:
the superconducting transition temperature and the magnetic
ordering temperature for the $^{16}$O/$^{18}$O substituted
samples, respectively, $\alpha_{T_c}=-{\rm d}\ln T_c/{\rm d}\ln
M_{\rm O}$: the OIE exponent of $T_c$, $\alpha_{T_m}=-{\rm d}\ln
T_m/{\rm d}\ln M_{\rm O}$: the OIE exponent of $T_m$. }
\begin{center}
 \vspace{-0.5cm}
\begin{tabular}{ccclllccc}\\
 \hline
 \hline
$x$&$^{16}T_c$&$^{18}T_c$&$\ \ \alpha_{T_c}$&$^{16}T_m$&$^{18}T_m$&$\alpha_{T_m}$\\
 &(K)&(K)&&\ (K)&\ (K)&\\
\hline
0.00&91.19(5)&90.99(4)&0.018(5)&\ \ --&\ \ --&--\\
0.20 &74.07(2)&73.27(2)&0.086(3)&\ \ --&\ \ --&--\\
0.30 &57.97(8)&56.79(7)&0.163(15)&1.13(11)&1.66(12)&-3.8(1.2)\\
0.40&44.80(2)&43.25(3)&0.277(7)&7.07(9)&7.56(9)&-0.55(31)\\
0.45&36.50(6)&35.12(6)&0.302(19)&15.54(13)&16.58(11)&-0.54(9)\\
0.50 &23.12(4)&20.16(4)&1.024(21)&17.82(11)&18.46(12)&-0.29(7)\\
0.55&14.4(2)&$<1.7$&7(1)\footnotemark[1]&21.05(18)&21.65(21)&-0.23(9)\\
0.58 &--&--&\ \ --&22.8(2)   &23.2(2)   &-0.13(9)\\
0.65 &--&--&\ \ --&32.3(4)   &34.4(4)   &-0.50(14)\\
0.70 &--&--&\ \ --&100.5(1.4)&116.8(1.3)&-1.30(14)\\
0.80 &--&--&\ \ --&210.2(4)  &212.6(4)  &-0.09(2)\\
1.00 &--&--&\ \ --&283.2(7)  &282.5(7)  &0.02(3)\\
\hline
0.00&91.35(4)\footnotemark[2]&91.16(4)\footnotemark[3]&0.017(5)&\ \ --&\ \ --&\ \ --\\
0.40&44.63(3)\footnotemark[2]&--&0.247(8)&7.10(9)&\ \ --&-0.52(30)\\

 \hline \hline \\

\end{tabular}
   \end{center}
 \vspace{-0.5cm}
  $^{a}${\small Estimated value (see text)  }
 \\
  $^{b}${\small Back exchanged $^{18}$O$\rightarrow$$^{16}$O sample }
 \\
 $^{c}${\small Back exchanged $^{16}$O$\rightarrow$$^{18}$O sample}
 \vspace{-0.3cm}

\end{table*}
The results of the OIE's on $T_c$, $T_g$ and $T_N$ are summarized
in Table~\ref{Table:OIE-results} and
Fig.~\ref{fig:Phase-Diagramm}. The doping dependences of $T_c$,
$T_g$, and $T_N$ for the $^{16}$O substituted samples are in
agreement with the results of Cooke {\it et al.} \cite{Cooke90}.
The second magnetic transition at $T_{N_2}\simeq17$~K (observed
for $x= 0.7\div1.0$) which is associated with the ordering of the
Pr sublattice is not considered here.

\begin{figure}[htb]
\includegraphics[width=0.8\linewidth, angle=90]{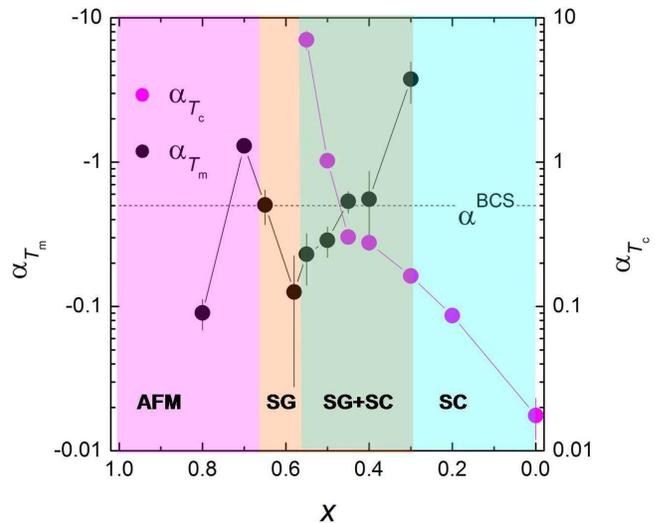}
 \vspace{-0.3cm}
\caption{OIE exponents $\alpha_{T_c}$ and $\alpha_{T_m}$ ($m=g$,
$N$) for $^{16}$O/$^{18}$O substituted
Y$_{x}$Pr$_{1-x}$Ba$_2$Cu$_3$O$_{7-\delta}$ as a function of the Pr
content $x$. The dashed line corresponds to $\alpha^{\rm
BCS}_{T_c}=0.5$. The meaning of the areas denoted by ''AFM``,
''SG``, ''SG+SC``, and ''SC`` are the same as in
Fig.~\ref{fig:Phase-Diagramm}.}
 \label{fig:alpha-beta}
\end{figure}

It is convenient to quantify the OIE's on the  transition
temperatures $T_y$ ($y$ denotes $c$, $g$, or $N$)  in terms of the
isotope effect exponents defined by:
\begin{equation}
\alpha_{T_y}= -\frac{d \ln T_y}{d \ln M_{\rm O}} = - \frac{\Delta
T_y/T_y}{\Delta M_{\rm O}/M_{\rm O}}= - \frac{(^{18}T_y-\
^{16}T_y)/^{16}T_y}{(^{18}M_{\rm O}-\ ^{16}M_{\rm O})/^{16}M_{\rm
O}},
 \label{eq:BCS-alpha}
\end{equation}
where $M_{\rm O}$ is the mass of the oxygen isotope
($^{16}$O/$^{18}$O). The values of $\alpha_{T_c}$ and
$\alpha_{T_m}$ ($m=g$, $N$) are listed in
Table~\ref{Table:OIE-results} and shown in
Fig.~\ref{fig:alpha-beta} as a function of Pr content $x$. For
$0.0\leq x\leq0.5$ the values of $\alpha_{T_c}$ are in agreement
with previous results \cite{Franck91Franck94}. In order to
estimate $\alpha_{T_c}$ for $x=0.55$, we assume for $^{18}T_c$ the
conservative value $^{18}T_c=1.7(1.7)$~K, yielding
$\alpha_{T_c}=7(1)$. In the SG phase a high value of
$\alpha_{T_m}=-3.8(1.2)$ for $x=0.3$ was found, in accordance with
a previous study of the SG behavior in Mn doped
La$_{2-x}$Sr$_x$CuO$_4$ \cite{Shengelaya99}.
Both $\alpha_{T_c}$ and $\alpha_{T_m}$ exhibit unusual features,
{\it i.e.}:
(i) $\alpha_{T_c}$ and $\alpha_{T_m} (m = g, N)$ depend  strongly on
$x$, being small in amplitude at ''extreme`` Pr content ($x=0.0$ for
$T_c$ and $x=1.0$ for $T_N$) and strongly increase upon approaching
$x=0.55$ and $x=0.3$, respectively (see Fig.~\ref{fig:alpha-beta}).
For $0.5\leq x\leq0.55$, $\alpha_{T_c}$ exceeds considerably the BCS
isotope-exponent $\alpha^{\rm BCS}_{T_c}=0.5$.
(ii) $\alpha_{T_c}$ increases monotonically with increasing $x$,
while $\alpha_{T_m}$ has a pronounced maximum around $x\simeq0.7$.
This maximum is a consequence of the fact that $T_N$ depends much
stronger on $x$ than $T_g$ (see Fig.~\ref{fig:Phase-Diagramm}).
(iii) $\alpha_{T_c}$ and $\alpha_{T_m}$ have {\it opposite} sign,
{\it i.e.}, $T_c$ decreases with increasing oxygen-isotope mass
($^{16}T_c>$$^{18}T_c$), whereas $T_g$ and $T_N$ increases
($^{16}T_g<$$^{18}T_g$, $^{16}T_N<$$^{18}T_N$, except for
$x=1.0$). This is particulary interesting in the region of the
phase diagram where superconductivity (SC) and the spin-glass (SG)
magnetism coexist (see Figs.~\ref{fig:Phase-Diagramm} and
\ref{fig:alpha-beta}).
(iv) The strong increase of both $\alpha_{T_c}$ and $\alpha_{T_m}$
by approaching $x=0.55$ and $x=0.3$,  respectively (see
Fig.~\ref{fig:alpha-beta}) suggests that the critical
concentrations where superconductivity ($x^{crit}_{T_c}$) and
magnetism ($x^{crit}_{T_m}$) disappear are different for $^{16}$O
and $^{18}$O substituted samples. An analysis of  the
$^{16}$O/$^{18}$O data for $T_c(x)$ presented in
Fig.~\ref{fig:Phase-Diagramm} by means of the power law
$T_c(x)=T_c(x=0)[1-(x/x^{crit}_{T_c})^{\delta}]^{\beta}$ yields:
$^{16}x^{crit}_{T_c}=0.570(1)$, $^{16}\delta=1.33(4)$,
$^{16}\beta=0.72(3)$ and $^{18}x^{crit}_{T_c}=0.556(2)$,
$^{18}\delta=1.33(2)$, $^{18}\beta=0.73(3)$. A linear
extrapolation of $^{16}T_g(x)$ and $^{18}T_g(x)$ in the region
$x\simeq0.3\div0.5$ to $T_g=0$ yields:
$^{16}x^{crit}_{T_m}\simeq0.287(2)$ and
$^{18}x^{crit}_{T_m}\simeq0.278(2)$. It is interesting to note
that the relative oxygen-isotope shifts of the critical
concentrations $x^{crit}_{T_c}$ and  $x^{crit}_{T_m}$, defined by
$\Delta x/x=(^{18}x-$$^{16}x)/^{16}x$, are the same within
experimental error: $\Delta x^{crit}_{T_c}/x^{crit}_{T_c} = -
2.5(4)$\% and $ \Delta x^{crit}_{T_m}/ x^{crit}_{T_m} = - 3.1(7)
$\%.

Recently, Ofer {\it et al.} \cite{Ofer06} studied a series of
Ca$_x$La$_{1-x}$Ba$_{1.75-x}$La$_{0.25+x}$Cu$_3$O$_y$ HTS's with
various doping levels -- from undoped to highly overdoped. They
concluded that all characteristic temperatures $T_N$, $T_g$, and
$T_c$ are controlled by the in-plane magnetic exchange energy $J$
of the undoped (parent) compound, thus inferring that the isotope
effects on $T_N$, $T_g$, and $T_c$ are driven by the isotope
dependence of $J$. In order to get a rough estimate of the OIE on
$J$ we employed the method proposed by Zhao {\it et al.}
\cite{Zhao94,Zhao01}. For the undoped antiferromagnetic parent
compounds the following relation holds: $\Delta T_N/T_N \simeq
\Delta J/J \cdot B/(B+1)$, where $B=2J/T_N\simeq10$. For $x=1.0$
with $^{16}T_N=283.2(7)$ and $^{18}T_N=282.5(7)$ (see
Table~\ref{Table:OIE-results}) one obtains $\Delta J/J \simeq
0.3(4)$\%, {\it i.e.}, the OIE on $J$ is zero within experimental
uncertainty. Therefore, $J$ is very unlikely  the source of the
observed OIE's on $T_N$, $T_g$, and $T_c$.

Opposite to the above interpretation, we argue that the observed
OIE's on $T_g$ and $T_N$ are directly related to the different
charge carrier mobilities (renormalized kinetic energies
\cite{Bussmann-Holder07}) in the $^{16}$O/$^{18}$O substituted
samples caused by the isotope dependence of the charge carrier
mass $m^\ast$. As shown in
Refs.~\cite{Zhao97Zhao98,Hofer00,Zhao01,
Khasanov04a,Khasanov04Khasanov06Khasanov07} the charge carriers in
the $^{18}$O substituted samples are heavier and, consequently,
less mobile than in $^{16}$O substituted samples. The recent
studies of H\"ucker {\it et al.} \cite{Hucker99Hucker04} clearly
demonstrate that increasing the hole mobility in
La$_{2-x}$Sr$_x$CuO$_4$ rapidly suppresses the antiferromagnetic
ordering temperature $T_N$. A similar conclusion was also reached
by Shengelaya {\it et al.} \cite{Shengelaya99} based on the
observation of a giant OIE on the spin-glass ordering temperature
$T_g$ in Mn doped La$_{2-x}$Sr$_x$CuO$_4$.

As for OIE on $T_c$, the observation of the onset of
superconductivity at different $x^{crit}_{T_c}$ (see
Fig.~\ref{fig:Phase-Diagramm}) implies that the number of carriers
condensed into Cooper pairs are larger for $^{16}$O substituted
samples then for $^{18}$O ones. Consequently, we propose that the
different hole mobilities in $^{16}$O/$^{18}$O substituted samples
are also responsible for the oxygen isotope shift of the critical
concentration $x^{crit}_{T_c}$ and then necessarily also for the
OIE on $T_c$. This statement is further supported by the following
facts: (i) The relative isotope shifts of $x^{crit}_{T_c}$ and
$x^{crit}_{T_m}$ are roughly the same $\Delta
x^{crit}_{T_c}/x^{crit}_{T_c}\simeq \Delta x^{crit}_{T_m}/
x^{crit}_{T_m}$ (see above), indicating that the OIE's on $T_c$,
$T_g$, and $T_N$ are of similar origin. (ii) In the region where
superconductivity and spin-glass magnetism coexist (see
Fig.~\ref{fig:Phase-Diagramm} and Table~\ref{Table:OIE-results})
the increase of $T_g$ in the $^{18}$O substituted sample (decrease
of the hole mobility) is associated with a corresponding decrease
of $T_c$.

In conclusion, oxygen isotope ($^{16}$O/$^{18}$O) effects on the
superconducting transition ($T_c$), the spin-glass ordering
($T_g$), and the antiferromagnetic ordering ($T_N$) temperatures
were studied for a series of
Y$_{1-x}$Pr$_x$Ba$_2$Cu$_3$O$_{7-\delta}$ samples as a function of
Pr content ($0.0\leq x\leq1.0$). The OIE exponent $\alpha_{T_c}$
increases with increasing $x$ (decreasing doping) reaching a
maximum at $x^{crit}_{T_c} \approx 0.55$, where superconductivity
disappears. At $x=1.0$ (undoped case) $\alpha_{T_N} \simeq 0$
within experimental uncertainty. For decreasing $x$ (increasing
doping) $\alpha_{T_m}$ increases, reaching a maximum at
$x^{crit}_{T_m} \approx 0.3$ where spin-glass behavior vanishes.
In the range of $0.3 \leq  x < 1.0$ the OIE's on $T_g$ and $T_N$
are {\it sign reversed} as compared to the one on $T_c$. The
relative isotope shift of the critical Pr concentration where
superconductivity ($x^{crit}_{T_c}$) and magnetism
($x^{crit}_{T_m}$) disappear are the same $\Delta
x^{crit}_{T_c}/x^{crit}_{T_c}\simeq \Delta x^{crit}_{T_m}/
x^{crit}_{T_m}$, indicating that the OIE's on $T_c$, $T_g$, and
$T_N$ are interrelated. These OIE's are suggested to arise from
the isotope dependent mobility of the charge carriers as proposed
in a model where polaronic renormalization of the single particle
energies are introduced \cite{Bussmann-Holder07}. The formation of
polaronic charge carriers may be caused by a strong Jahn-Teller
effect \cite{Bednorz86}, in close analogy to doped perovskite
manganites \cite{Zhao96}.
The unconventional isotope effects presented here clearly
demonstrate that lattice effects play a significant role in the
physics of cuprates in both the magnetic and the superconducting
state.

This work was partly performed at the Swiss Muon Source (S$\mu$S),
Paul Scherrer Institute (PSI, Switzerland). The authors are grateful
to A.~Bussmann-Holder and K.~Alex~M\"uller  for many stimulating
discussions, A.~Amato, R.~Scheuermann, and D.~Herlach for providing
the instrumental support during the $\mu$SR experiments. This work
was supported by the Swiss National Science Foundation, by the
K.~Alex M\"uller Foundation and in part by the SCOPES grant No.
IB7420-110784, the EU Project CoMePhS, and the NCCR program MaNEP.

\end{document}